\documentclass[twocolumn]{revtex4}
\usepackage{graphicx}
\usepackage{amsmath,amsthm}
\begin{document}

\title{Structural properties of Pb$_{3}$Mn$_{7}$O$_{15}$ determined from high-resolution synchrotron powder diffraction}

\author{J.C.E. Rasch$^{1,2}$}
\email{rasch@ill.fr}
\author{D.V. Sheptyakov$^2$}
\author{J. Schefer$^2$}
\author{L. Keller$^2$}
\author{M. B\"ohm$^1$}
\author{F. Gozzo$^3$}   
\author{N.V. Volkov$^4$}
\author{K.A. Sablina$^4$}
\author{G.A. Petrakovskii$^4$}
\author{H. Grimmer$^5$}
\author{K. Conder$^5$}
\author{J.F. L\"offler$^6$}

\affiliation{$^1$Institut Laue-Langevin, 6 rue Jules Horowitz, BP 156, 38042 Grenoble Cedex 9, France}
\affiliation{$^2$Laboratory for Neutron Scattering, ETH Zurich \& Paul Scherrer Institut, CH-5232 Villigen, PSI, Switzerland}
\affiliation{$^3$Swiss Light Source, Paul Scherrer Institut, CH-5232 Villigen, PSI, Switzerland}
\affiliation{$^4$L.V. Kirensky Institute of Physics, SB RAS, Krasnoyarsk 660036, Russia}
\affiliation{$^5$Laboratory for Developments and Methods, Paul Scherrer Institut, CH-5232 Villigen, PSI, Switzerland}
\affiliation{$^6$Laboratory of Metal Physics and Technology, Department of Materials, ETH Zurich, 8083 Zurich, Switzerland}

\date{\today}

\begin{abstract}
We report on the crystallographic structure of the layered compound Pb$_{3}$Mn$_{7}$O$_{15}$. Previous analysis based on laboratory X-ray data at room 
temperature gave contradictory results in terms of the description of the unit cell. Motivated by recent magnetic bulk measurements of this system~ 
\cite{Volkov2008a}, we re-investigated the chemical structure with high-resolution synchrotron powder diffraction at temperatures between 15~K and 
295~K. Our results show that the crystal structure of stoichiometric Pb$_{3}$Mn$_{7}$O$_{15}$ has a pronounced 2-dimensional character and can be 
described in the orthorhombic space group \textit{Pnma}.
\end{abstract}

\maketitle

\section{Introduction}

Since the discovery of the colossal magneto-resistance (CMR) effect in (La,Pb)MnO$_3$ \cite{Searle1969}, manganites have been comprehensively 
investigated throughout the last years, e.g. Chatterji \cite{Chatterji2004}. Extensive studies focused especially on members of two groups: the 
perovskite-type compounds (space groups $R\bar{3}m$, $Pbnm$) and the layered compounds of the so-called Ruddlesden-Popper series (space group 
$I/4mmm$) with the general formula A$_{n+1}$Mn$_n$O$_{3n+1}$, where A is a rare-earth metal and $n$ is the number of corner-shared MnO$_6$ octahedral 
sheets forming the layer~\cite{Chatterji2004,Tokura2000}.\\
Pb$_{3}$Mn$_{7}$O$_{15}$ cannot be classified in either of these groups, although, like the $n=1$ compounds of the Ruddlesden-Popper series, it grows 
in a pronounced crystalline anisotropy with alternating monolayers of MnO$_6$ octahedra and PbO sheets. The first synthesis was reported in literature 
already thirty years ago~\cite{Darriet1978}, but until recent times little was known about the physical properties. No evidence of CMR is present in 
the pure compound. Nevertheless, the layered type structure and the presence of a mixed valence state of Mn ions (Mn$^{3+}$/Mn$^{4+}$) recently 
motivated a new thorough investigation:
Magnetisation and specific heat measurements on Pb$_{3}$Mn$_{7}$O$_{15}$ revealed several magnetic phases including long-range magnetic ordering below 
$T_N=70$~K \cite{Volkov2008a} and paramagnetic charge localisation at $T_{CL}=250$~K. Dielectric properties give rise to formation of polarons, which 
make Pb$_{3}$Mn$_{7}$O$_{15}$ a promising candidate for multiferroic materials \cite{Volkov2008b}.\\ 
However, the basis for further microscopic investigations of the physical properties is the exact knowledge of the crystal symmetry. The structural 
properties of the layered compound Pb$_{3}$Mn$_{7}$O$_{15}$ have been a matter of debate since the first report on its structure by Darriet \textit{et 
al.} \cite{Darriet1978}, using single crystal X-ray diffraction data taken with a laboratory K$_{\alpha}$ molybdenum source. The crystallographic 
structure was refined based on the orthorhombic space group $Cmc2_1$. Marsh \& Herbstein \cite{Marsh1983} reformulated the structure based on the data 
of Darriet \textit{et al.}~\cite{Darriet1978} in the orthorhombic space group $Cmcm$ which involved shifting the origin by about 0.25 in 
$z$-direction. Symmetrizing the positions of some pairs of atoms by shifts up to 0.2~\AA\:was thereby necessary. Shortly afterwards, Le Page \& 
Calvert~\cite{LePage1984} found that a description in $P6_3/mcm$ discloses a new class of systematic absences not accounted for previously. The new 
hexagonal unit cell of Pb$_{3}$Mn$_{7}$O$_{15}$ is related to the orthorhombic description in $Cmcm$ by 
$\mathbf{a_\mathrm{hex}}=(\mathbf{a}_\mathrm{ortho}-\mathbf{b}_\mathrm{ortho})/2,\;
\mathbf{b}_\mathrm{hex}=\mathbf{b}_\mathrm{ortho},\; \mathbf{c}_\mathrm{hex}=\mathbf{c}_\mathrm{ortho}$.
Nevertheless, Weissenberg and Laue patterns also indicate contradictory results concerning hexagonal and orthorhombic symmetry \cite{Darriet1978}. In 
order to settle this discrepancy between the different models and in order to investigate the structural properties as a function of temperature, we 
performed new measurements on this material employing high-resolution synchrotron radiation. 

\section{Experimental Details}
\subsection{Sample Preparation and Characterisation}

Single crystals were grown by the flux method, as described in \cite{Volkov2008a}. The flux agent PbO was chosen since it is known as an effective 
solvent for many oxide compounds and allows to avoid incorporation of foreign ions into the lattice. The synthesis of Pb$_{3}$Mn$_{7}$O$_{15}$ single 
crystals started with heating a mixture of 93 \% by weight of high purity PbO and 7~\% by weight of Mn$_2$O$_3$ in a platinum crucible at 
1000~$^{\circ}$C for 4~h. Afterwards, the crucible was cooled to 900~$^{\circ}$C at a cooling rate of $2.5$~$^{\circ}$C/h, followed by non-constrained 
cooling to room temperature. Single crystals with a plate-like hexagonal form and a shiny black surface were found on the solidified liquid surface. 
The plates measured up to 40~mm in diameter and were extracted mechanically from the flux. Afterwards, the crystals were carefully ground for 
synchrotron powder diffraction investigations.\\
Thermogravimetric analysis was employed to verify the oxygen content in our sample. The powder sample was heated from room temperature up to 
800~$^{\circ}$C in a 5$\%$ H$_2$/ 95$\%$ He flux at a heating rate of 2 $^{\circ}$C/ minute. A mass spectrometer analysed the evolution of H$_2$O. 
Finally, from the weight loss of the sample the initial oxygen content was calculated. The oxygen content of our sample has been determined to 
Pb$_{3}$Mn$_{7}$O$_{x}$ with $x=14.93\pm 0.05$.

\subsection{High-resolution synchrotron measurements}

High-resolution synchrotron radiation powder diffraction patterns were collected at the powder diffraction station of the Swiss Light Source Materials 
Science (SLS-MS) beamline using a multi crystal analyzer detector \cite{Hodeau1998,Gozzo2004}. The instrumental contribution to the peak profile was 
reduced to negligible levels by carefully optimizing the beamline optics and the detector while observing the crystal analyzer rocking curve with the 
attenuated direct beam \cite{Gozzo2007}. The photon energy ($\lambda=0.617793(1)$~\AA) was determined using the silicon standard from NIST Si640c, 
whereas the instrumental resolution function was evaluated using Na$_2$Ca$_3$Al$_2$F$_{14}$, a standard powder known for being characterized by Bragg 
peaks with a negligible intrinsic contribution to the instrumental line shape.
\begin{figure}
\includegraphics[width=0.5\textwidth]{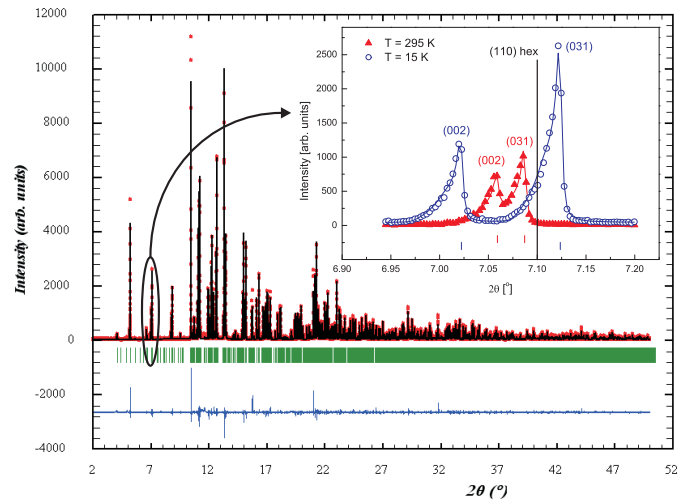}
\caption{Observed and calculated neutron diffraction powder pattern of Pb$_{3}$Mn$_{7}$O$_{15}$ at $T=15$~K.   
The insert shows the splitting of the hexagonal (110) reflection into orthorhombic (002) and (031) at $T=15$~K (open circles, blue online) and 
$T=295$~K (closed triangles, red online). 
Please note the interchange of the hexagonal $c$-axis with the orthorhombic $a$-axis.}
\label{fig:15k-ref}
\end{figure}
During the acquisitions, the detector arm was continuously rotated at a constant speed, recording the data at high reading frequency, which are 
rebinned to the appropriate step size after the measurements. Temperature dependent measurements between 15~K and 295~K in steps of 50~K were 
performed using the Janis Cryostat. The Pb$_{3}$Mn$_{7}$O$_{15}$ powder was mounted in the cryostat in 0.2~mm Lindemann capillaries and the capillary 
spun at approximately 10~Hz during the 2$\theta$ scan.

\section{Results}

\begin{figure}
\includegraphics[width=8cm,angle=0]{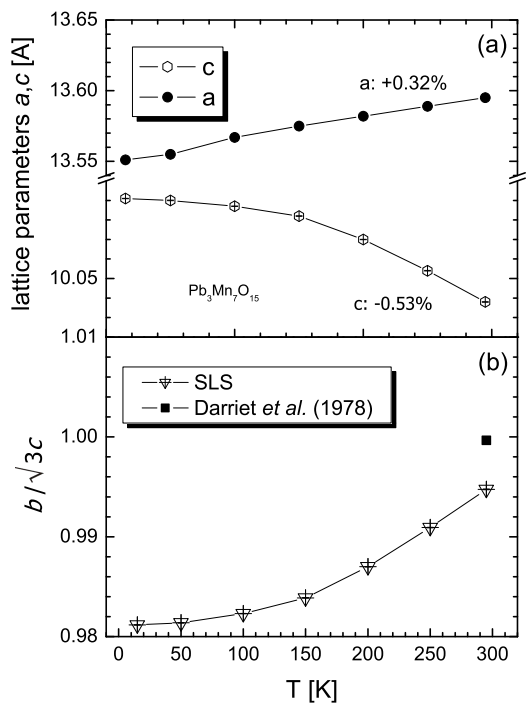}
\caption{(a) Temperature dependence of the lattice parameters $a$ and $c$ in Pb$_{3}$Mn$_{7}$O$_{15}$. The orthogonal $a$-axis is vertical to the 
layer structure and corresponds to the hexagonal $c$-axis in the description of \cite{LePage1984}. (b) The in-plane lattice parameter ratio 
corresponding to the hexagonal lattice (black square) is not reached up to 295~K, therefore only an orthorhombic description (open triangles) is 
possible in our case.}
\label{fig:cell-parameters}
\end{figure}

The powder diffraction pattern at $T$=15~K is shown in Fig.~\ref{fig:15k-ref}.\\
Refinements, based on different space groups reported earlier \cite{Darriet1978,Marsh1983,LePage1984}, gave unsatisfactory agreement with the observed 
pattern, due to a systematic splitting of peaks indicating a lowering of the space group symmetry. The inset of Fig.~\ref{fig:15k-ref} shows a small 
angular section around the expected (110) reflection in a hexagonal setting as reported by Darriet \textit{et al.} \cite{Darriet1978}.
The splitting was observed at all temperatures, excluding a structural phase transition as shown by the continuous change of the orthogonal 
distortion, not reaching the hexagonal symmetry up to 295~K (Fig. \ref{fig:cell-parameters}b). The value of splitting shows a temperature dependence 
and amounts to $2\theta=0.03^{\circ}$ and $T=295$~K to $2\theta=0.1^{\circ}$ at $T=15$~K. From an investigation of the subgroup relationships starting 
from the high symmetry hexagonal $P6_3/mcm$ structure, the orthorhombic space group $Pnma$ is obtained. The different structures reported earlier are 
related in the following way: Orthorhombic $Cmcm$ used by Marsh \& Herbstein \cite{Marsh1983} is a \textit{translationsgleiche} (a symmetry element is 
removed, but translation symmetry is maintained) subgroup of $P6_3/mcm$ of index~3. $Cmc2_1$ used by Darriet \textit{et al.} \cite{Darriet1978} is 
also \textit{translationsgleich} of index 2 of space group \textit{Cmcm}.
Space group \textit{Pnma} is reached by a \textit{klassengleiche} (in our case, the C-centering has been removed) subgroup of \textit{Cmcm} of index 
2. The crystallographic $a$-axis is now perpendicular to the layers, in contrast to previous descriptions with perpendicular $c$-axis.\\
All preceding interpretations were based on the same data set taken by Darriet \textit{et al.} \cite{Darriet1978} on a laboratory X-ray single crystal 
diffractometer. Although the angular resolution at the synchrotron is considerably improved compared to a laboratory instrument, the peak splitting of 
about $2\theta=0.03^{\circ}$ at room temperature is resolvable on a standard X-ray laboratory instrument. A possible explanation for the discrepancy 
in the structural description might be a different oxygen content of our sample compared to the previous studies. Investigations on the layered 
manganites BaTb$_2$Mn$_2$O$_7$ and BaSm$_2$Mn$_2$O$_7$ \cite{kamegashira1,kamegashira2} have shown that phase transitions from tetragonal to 
orthorhombic symmetry occur as a function of oxygen or nitrogen annealing, suggesting some tolerance for oxygen non-stoichiometry.\\
\begin{figure}
\includegraphics[width=8cm]{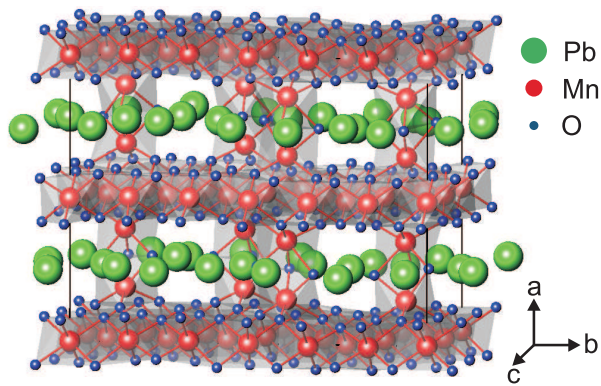}
\caption{Layered structure of Pb$_{3}$Mn$_{7}$O$_{15}$. The Pb atoms are interconnecting the
Mn-O-layers (on-line version: Mn (red); O (blue); Pb (green)).}
\label{fig:Pb3Mn7O15-layered-structure}
\end{figure}
Based on the $Pnma$ space group we obtained satisfactory agreement between the measured data sets and calculations ($R=8.9~\%$ for $T=15$~K and 
$R=13.8~\%$ for $T=295$~K) as shown in Table~\ref{tbl:ref-comparison}. The unit cell comprises 8 formula units leading to 30 independent atoms per 
unit cell (see Fig.~\ref{fig:Pb3Mn7O15-layered-structure}). Manganese ions are distributed among the general 8$d$ position and the special positions 
4$a$, 4$b$ and 4$c$. The MnO$_6$ octahedra can be grouped into undistorted (4$a$, 4$b$), weakly distorted (4$c$, 8$d$) and strongly distorted 
octahedra (8$d$) with respect to the angles of the main axis of the octahedra (see Table~\ref{tbl:octahedras}). The weakly and undistorted octahedra 
form layers around $x=0$ and $x=0.5$ bridged by pairs of strongly distorted octahedra. The layers show a small modulation within the unit cell along 
the $b$ direction, where the mean displacement along the $a$ direction of all manganese atoms within a layer decreases with temperature from 
$\Delta\bar{x}=0.005$ relative length units at $T=295$~K  to $\Delta\bar{x}=0.003$ at $T=15$~K. The decrease of the modulations with temperature goes 
with a reduction of the lattice constant $b$ with simultaneous increase along the $c$ direction, so that the pseudo-hexagonal arrangement of Mn atoms 
gets stretched along $c$ at low temperatures. The variation of lattice parameters as function of temperature is shown in 
Fig~\ref{fig:cell-parameters}.
\begin{figure}
\includegraphics[width=8cm,angle=0]{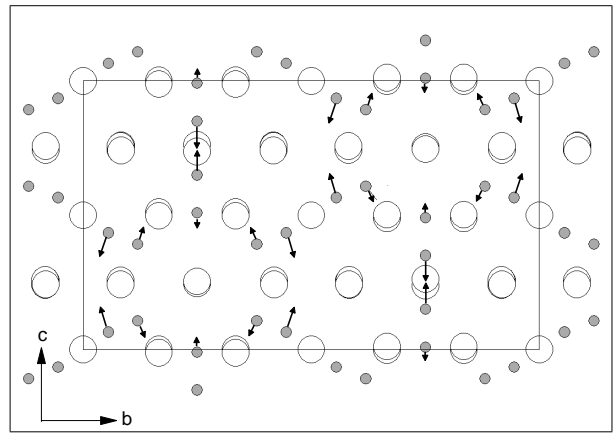}
\caption{Projection of the low temperature structure ($T=15$~K) on the $b$-$c$ plane. Only manganese (white open circles) and lead (gray closed 
circles) are shown.
Arrows indicate the direction of displacements of the lead atoms from $T=295$~K to $T=15$~K.}
\label{fig:leadatoms15K}
\end{figure}\\
As reported earlier~\cite{Darriet1978}, the Pb atoms, equally located on two different symmetry sites $8d$ and $4c$, are closely confined around 
layers at $x=0.25$ and $x=0.75$, half-way in-between the MnO$_6$ layers. The Pb layers follow a similar modulation along $b$ with displacement along 
$a$ as the MnO$_6$ layers. Projected on the (100) planes the lead atoms lie on lines along the $<$011$>$ and $<$001$>$ directions. At $T=15$~K the 
lead atoms get displaced as indicated in Fig.~\ref{fig:leadatoms15K}, where the atoms of the $x=0.25$ layers are systematically shifted towards the 
positive $c$ directions, while atoms on the $x=0.75$ layer are shifted in the opposite direction.

\section{Discussion}

As discussed by Volkov \textit{et al.} \cite{Volkov2008a} there are two factors which can cause polyhedral distortions of the MnO$_6$ octahedra in 
Pb$_{3}$Mn$_{7}$O$_{15}$. The first factor is the Jahn-Teller effect which originates from degenerate orbital states of the Mn$^{3+}$ ions in a 
regular octahedral crystal field. The valence state of the Mn ions is discussed explicitly below. Secondly the stereo-active lone pairs of the 6s$^2$ 
electrons in Pb$^{2+}$ can lead to significant shifts of some oxygen positions due to repulsion of the lone pair with Pb-O bonds. This assumption can 
be supported by the fact, that strongly distorted MnO$_6$ octahedra consist of oxygen ions which are in close proximity to Pb ions.\\
According to the stoichiometry of the formula unit, the Mn ions in Pb$_{3}$Mn$_{7}$O$_{15}$ are in a mixed valence state Mn$^{3+}$/Mn$^{4+}$ with a 
ratio 4/3. Assuming the Jahn-Teller effect to be responsible for the distortion, one can distribute Mn$^{3+}$ ions onto interlayer Mn sites (Mn4, Mn5) 
and Mn$^{4+}$ ions onto the undistorted positions (Mn6, Mn7) in a straightforward way. Assuming further that the remaining Mn atoms have equal 
probability to be in either of the oxidation state, one obtains the total ratio of Mn$^{3+}$/Mn$^{4+}=4/3$ corresponding to the stoichiometry of the 
formula unit. The distribution of tri- and tetravalent Mn ions can be estimated from Mn-O bond lengths via bond-valence sums (BVS). The BVS is 
proportional to the sum over the deviation $(r_0-r)$, where $r$ is the measured bond length and $r_0$ an emperical parameter for the mean cation-anion 
distance. Since the $r_0$ values for Mn$^{3+}$ and Mn$^{4+}$ are very similar, one single $r_0$ value (1.76~\AA) can be used as an estimate for the 
calculations \cite{Brown1985}. The result (Table \ref{tbl:octahedras}) is less satisfactory than reported by Volkov \textit{et al.} 
\cite{Volkov2008a}, but we assume it more accurate since the lower symmetric space group we used allows an investigation of non-averaged bond-lengths. 
After this analysis it is less evident to connect unambiguously the strongly distorted octahedra (Mn4, Mn5) with the Jahn-Teller effect (see Table 
\ref{tbl:octahedras}). At this point no final conclusion can be drawn about the mixed valence state in Pb$_{3}$Mn$_{7}$O$_{15}$ since Mn ions on all 
crystallographic sites seem to be in average in a state between tri- and tetravalency which cannot be connected to the degree of octahedral 
distortion.

\section{Conclusions}

We re-investigated the mixed valence Mn$^{3+}$/Mn$^{4+}$ manganite Pb$_{3}$Mn$_{7}$O$_{14.93}$ using X-ray synchrotron powder diffraction. The oxygen 
content was verified by thermogravimetry. Rietveld refinement showed that our data can be well described within the orthorhombic space group $Pnma$ 
and that there is no structural phase transition in the observed temperature range 15 $\le$ $T$ $\le$ 295 K. The refined space group is different to 
several previous descriptions reported in literature, which were all analysed from the data set taken by Darriet \textit{et al.} \cite{Darriet1978}. 
The crystal symmetry stays unchanged in the investigated temperature regime, the pseudo-hexagonal arrangement of the Mn atoms gets stretched along the 
$c$ axis at low temperatures and the Pb atoms experience a systematic displacement. Earlier investigations \cite{Volkov2008a} based on a higher 
symmetry phase and magnetization data suggest three different types of MnO$_6$ octahedra distinguished by the degree of distortion and the valence 
state of the Mn ions. Our synchrotron measurements on the same crystals as in \cite{Volkov2008a} show that this classification is less significant in 
the lower symmetric description and that BVS calculations could not clarify unambiguously the Mn valence states.

\section{Acknowledgement}
Synchrotron beam time at the Swiss Light Source Materials Science beamline Powder Diffraction Station is gratefully acknowledged as well as the 
support of the crystal growth department of the L.V. Kirensky Institute in Krasnoyarsk and the Laboratory for Developments and Methods (LDM) of the 
PSI. The work is supported by INTAS grant 06-1000013-9002 of the Russian Academy of Science (RAS), Siberian Branch.

\begin{table}
\caption{Comparsion of structural models of  Pb$_{3}$Mn$_{7}$O$_{15}$} 
\label{tbl:ref-comparison} {\smallskip}
\begin{small}
\begin{tabular}{lccccc}
\hline
Reference                                     & \cite{Darriet1978}    & \cite{Marsh1983}  & \cite{LePage1984}     & This paper & This paper\\
 Temperature  [K]                             &    295              &    295        &    295              & 295       & 15 \\
space group                                   & \textit{Cmc}2$_1$ (No. 36)   & \textit{Cmcm} (No. 63)  & \textit{P6$_3$/mcm} (No. 193) & \textit{Pnma} 
(No. 62) & \textit{Pnma} (No. 62)\\
                                              & orthorhombic        & orthorhombic  & hexagonal           & orthorhombic &  orthorhombic\\
\hline

cell parameters                            &                     &               &                     &              \\  
\hline
               a [\AA]                        & 17.28 (1)            & 17.28 (1)      & 9.98 (1)           &   13.595100 (21)   & 13.551275(15)\\
               b [\AA]                        & 9.98 (1)             &  9.98 (1)      & 9.98 (1)           &    17.295435 (25)  & 17.148983(18)\\
               c [\AA]                        & 13.55 (1)            & 13.55 (1)      & 13.55 (1)          &   10.038134 (14)   & 10.09088(10)\\
			   volume [\AA$^3$]               & 2336.76              & 2336.76        & 1168.78            &   2360.30          & 2345.03\\
data                                          & \cite{Darriet1978}   & \cite{Darriet1978}  & \cite{Darriet1978}     & SLS-MS & SLS-MS \\
sample                                        & single crystal      &   &      & polycrystalline &polycrystalline\\ 
wavelength     [\AA]                          & Mo K$_{\alpha}$      &   &      & 0.617793~(1) & 0.617793~(1) \\
number of reflections                         & 904                 &   &       &    n.a. & n.a.   \\
sin($\Theta$/$\lambda $) [\AA$^{-1}$]         & 0.67      &   &      & 0.65   &0.65 \\
$\chi ^2$ &   & & & 1.70 & 2.26 \\
R   [\%]      &   &  & & 13.8 & 8.9 \\
R$_\mathrm{Bragg}$ $[\%] $ & & & & 6.9 &  4.4 \\
\end{tabular}
\end{small} 
\end{table}

\begin{table}
\caption{O-Mn-O angles $\alpha$ and atomic distances $d(O-Mn)$, $d(Mn-O)$ of the MnO$_6$ octahedra in Pb$_3$Mn$_7$O$_{15}$ at $T=15$~K.} 
\label{tbl:octahedras} {\smallskip}
\begin{small}
\begin{tabular}{lcccccc}
Atom	& Site  & &   $\alpha$   &   $d(O-Mn)$    &   $d(Mn-O)$ & BVS  \\
\hline
 Mn1 & 8d & O1-Mn1-O12  & 172 (2)     & 1.97 (2) & 1.86 (2) & +3.458 (89)\\
     &    & O2-Mn1-O12  & 173 (2)     & 1.87 (2) & 2.06 (2)\\
     &    & O16-Mn1-O17 & 174.3 (1.8) & 2.16 (2) & 1.94 (2)\\
 Mn2 & 8d & O4-Mn2-O12  & 172.9 (1.9) & 1.92 (2) & 1.91 (2) & +3.700 (88)\\
     &    & O6-Mn2-O10  & 174.3 (2.0) & 2.00 (2) & 1.89 (2)\\
     &    & O9-Mn2-O16  & 173.7 (1.9) & 1.98 (2) & 1.95 (2)\\
 Mn3 & 8d & O5-Mn3-O13  & 173.8 (1.6) & 2.14 (2) & 2.28 (2) & +3.331 (79)\\
     &    & O7-Mn3-O11  & 171.5 (1.9) & 1.89 (2) & 2.04 (2)\\
     &    & O8-Mn3-O17  & 171.9 (2.0) & 1.82 (2) & 1.89 (2)\\
 Mn4 & 8d & O2-Mn4-O14  & 160.3 (1.6) & 2.13 (2) & 2.17 (2) & +3.121 (75)\\
     &    & O3-Mn4-O5   & 164.0 (1.8) & 1.96 (2) & 1.84 (2)\\
     &    & O10-Mn4-O15 & 161.6 (1.8) & 1.97 (2) & 2.05 (2)\\
 Mn5 & 8d & O1-Mn5-O14  & 166.9 (2.0) & 1.83 (2) & 2.00 (2) & +3.417 (89)\\
     &    & O3-Mn5-O4   & 161.8 (1.6) & 2.24 (2) & 2.13 (2)\\
     &    & O11-Mn5-O15 & 164.5 (1.9) & 1.81 (2) & 1.98 (2)\\
 Mn6 & 4a & O2-Mn6-O2   & 180.0 (1.9) & 1.95 (2) & 1.95 (2) & +3.385 (81)\\
     &    & O11-Mn6-O11 & 180 (2)     & 1.94 (2) & 1.94 (2)\\
     &    & O17-Mn6-O17 & 180.0 (1.9) & 2.03 (2) & 2.03 (2)\\
 Mn7 & 4b & O1-Mn7-O1   & 180 (2)     & 2.08 (3) & 2.08 (3) & +3.812 (101)\\
     &    & O10-Mn7-O10 & 180 (2)     & 2.00 (2) & 2.00 (2)\\
     &    & O16-Mn7-O16 & 180 (2)     & 1.77 (2) & 1.77 (2)\\
 Mn8 & 4c & O6-Mn8-O7   & 176 (3)     & 1.85 (3) & 2.09 (3) & +3.917 (113)\\
     &    & O12-Mn8-O13 & 175 (2)     & 1.92 (2) & 1.88 (2)\\
     &    & O6-Mn8-O7   & 175 (2)     & 1.92 (2) & 1.88 (2)\\
 Mn9 & 4c & O4-Mn9-O5   & 178.3 (1.9) & 1.88 (2) & 1.93 (2) & +4.034 (112)\\
     &    & O4-Mn9-O5   & 178.3 (1.9) & 1.88 (2) & 1.93 (2)\\
     &    & O8-Mn9-O9   & 175 (3)     & 2.04 (3) & 1.82 (3)\\
\end{tabular}
\end{small} 
\end{table}

\end{document}